\documentclass[12pt]{iopart}
\usepackage{graphicx,iopams}

\begin{document}

\title{Underlying mechanisms for normal heat transport in one-dimensional anharmonic oscillator systems with a double-well interparticle interaction}
\author{Daxing Xiong\footnote{Author to whom any correspondence should be addressed.}}
\address{Department of Physics, Fuzhou University, Fuzhou 350108, People's Republic of China}
\ead{phyxiongdx@fzu.edu.cn}

\begin{abstract}
Previous studies have suggested a crossover from superdiffusive to
normal heat transport in one-dimensional (1D) anharmonic oscillator
systems with a double-well type interatomic interaction like
$V(\xi)=-\xi^2/2+\xi^4/4$, when the system temperature is varied. In
order to better understand this unusual manner of thermal transport,
here we perform a direct dynamics simulation to examine how the
spreading processes of the three physical quantities, i.e., the
heat, the total energy and the momentum, would depend on
temperature. We find three main points that are worth noting: (i)
The crossover from superdiffusive to normal heat transport is well
verified from a new perspective of heat spread; (ii) The spreading
of the total energy is found to be very distinct from heat
diffusion, especially that under some temperature regimes, energy is
strongly localized, while heat can be superdiffusive. So one should
take care to derive a general connection between the heat conduction
and energy diffusion; (iii) In a narrow range of temperatures, the
spreading of momentum implies clear unusual non-ballistic behaviors;
however, such unusual transport of momentum cannot be directly
related to the normal transport of heat. An analysis of phonons
spectra suggests that one should also take the effects of phonons
softening into account. All of these results may provide insights
into establishing the connection between the macroscopic heat
transport and the underlying dynamics in 1D systems.
\end{abstract}

\pacs{05.60.-k, 44.10.+i}

\maketitle

\section{Introduction}
The viewpoint of anomalous heat transport in one-dimensional (1D)
momentum-conserving systems has now been widely
accepted~\cite{Report-1,Report-2}. The anomaly means that the
``standard" heat transport law, i.e., the Fourier's law of heat
conduction, stating that, the heat flux $\textbf{\emph{J}}$ is
proportional to the temperature gradient $\nabla T$:
$\emph{\textbf{J}}=-\kappa \nabla T$, with $\kappa$ the heat
conductivity assumed to be constant, is not validated. In
particular, for 1D anharmonic oscillator systems with conserved
momentum and symmetric interparticle interactions, it has now been
generally believed that $\kappa$ is not a constant but follows a
simple space $L$ scaling $\kappa \sim
L^{\alpha}$~\cite{Report-1,Report-2}. (For the momentum-conserving
systems with asymmetric interactions, refer to the recent
progress~\cite{Asymmetric-1,Asymmetric-2} and
debates~\cite{debate-1,debate-2,debate-3,debate-4}). Despite that
there is still no consensus on the universality classes of the
scaling exponent $\alpha$ and its accurate value(s), in most cases
$0 < \alpha <
1$~\cite{Ourwork-1,Ourwork-2,Ourwork-3,Universal-1,Universal-2,Universal-3,Universal-4,Universal-5}
can be concluded. This power-law space scaling has also been
corroborated by some relevant experimental studies of carbon
nanotubes~\cite{Carbon}.

Nevertheless, there are still at least two exceptional systems with
both conserved momentum and symmetric interactions against the above
belief~\cite{Note}, i.e., the 1D coupled rotator system and the
chain with a double-well (DW) potential. For both systems a
transition (or crossover) from superdiffusive to normal heat
transport (obeys the Fourier's law, $\alpha=0$) has been claimed to
take place. Several mechanisms have been proposed to understand the
observed normal transport in rotator systems. Early works related
the mechanism to the occurrence of phase jump~\cite{Rotator-1} and
the excitation of high-frequency stationary localized rotational
modes~\cite{Rotator-2}. While quite recently the mechanism has been
traced back to the diffusive behavior of the momentum
spread~\cite{YunyunLi} and the absence of the conserved quantity of
stretch from the perspective of nonlinear fluctuating
hydrodynamics~\cite{Rotator-3,Rotator-4}. However, when one turns to
the system with DW potential, the underlying mechanism for normal
heat transport has not yet been clarified so far. Even worse,
whether the transport would be normal or abnormal remains
controversial: early results indicated that heat conduction is
normal at low temperatures~\cite{Rotator-1}, nevertheless it was
doubted later by~\cite{Report-1,Report-2}. Two recent works
revisited the issue and suggested that normal behavior may appear in
a narrow temperature region near $T \simeq 0.1$~\cite{DW}. In a
quite recent work, we have also shown convincing evidences for this
normal behavior~\cite{Mywork}, however, the underlying mechanisms
are still not very clear.

In the present work we perform a further careful simulation to
examine the unusual heat transport and its underlying mechanisms in
the 1D oscillator systems with DW potential. For such purpose we
first consider the heat spreading process. To identify whether heat
transport is normal or anomalous, usually the space scaling
dynamical exponent $\alpha$ has been given the most efforts, for
which two kinds of dynamics simulation approaches, i.e., direct
nonequilibrium molecular dynamics simulations~\cite{Direct} and the
method based on Green-Kubo formula~\cite{GreenKubo}, have been
frequently used. However, just as raised by S. Olla~\cite{Workshop}
in a discussion session of a recent workshop: these studies for
deriving $\alpha$ usually ignored another key time scale. Therefore,
by viewing that concerning the time scaling may involve more
detailed information, which would enable us to present a more
detailed prediction for heat transport, here we shall investigate
the relaxation of equilibrium heat fluctuations of the system to
derive the space-time scaling properties to characterize the heat
transport behavior. We shall explore how this space-time scaling
depends on system temperature, through which we are able to verify
with satisfactory precision that, normal transport is indeed likely
to take place near the crossover temperature $T \simeq 0.1$.

To further understand this normal transport of heat, we shall also
investigate the relaxation of two other physical quantities
fluctuations, i.e., the total energy and the momentum. We will show
clear distinctions between the spreading of heat and the total
energy. In some temperature regimes, we will also reveal the unusual
non-ballistic behaviors of momentum spread. Nevertheless, the latter
features of momentum spread cannot be directly attributed to the
observed normal transport of heat. A careful analysis of the
system's phonons spectra indicates that, around the crossover
temperature point, phonons clearly tend to become softest. Based on
both facts, we thus conjecture that under the appropriate
temperature, phonons softening, together with the non-ballistic
behavior of momentum diffusion, may result in the observed normal
heat transport in the systems with DW potential.

\section{Model}
The considered model is a 1D momentum-conserving oscillator system
with Hamiltonian of the form
\begin{equation}
H=\sum_k \left[\frac{p_{i}^{2}}{2 \mu}+
V\left(q_{k+1}-q_{k}\right)\right],
\end{equation}
where $q_k$ denotes the displacement of the $k$-th particle from its
equilibrium position and $p_k$ its momentum. The mass $\mu$ is set
to be unity. The potential takes the DW type
\begin{equation}
V(\xi)=-\xi^2/2+\xi^4/4.
\end{equation}
\begin{figure}
\vskip-.2cm \hskip-0.4cm \centering
\includegraphics[width=10cm]{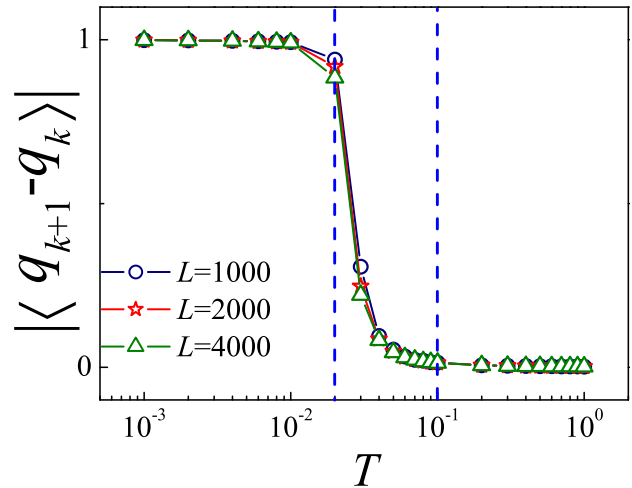} \vskip-0.3cm \caption{(Color
online) The order parameter $\left|\left< q_{k+1} - q_{k} \right>
\right|$ vs. temperature $T$, where the circles, stars and triangles
correspond to the results of space size $L=1000$, $2000$, $4000$;
and from left to right, the vertical dashed lines denote $T=0.02$
and $T=0.1$, respectively.}
\end{figure}

Such a system is very peculiar. First, equation (2) is an extension
of Fermi-Pasta-Ulam (FPU) interactions to the particular DW type,
which is usually adopted to model the structural phase
transition~\cite{PRL1975}. Interestingly, some recent works have
suggested that in the trapped-ion chains, heat transport can be
feasibly tuned across the structural phase
transition~\cite{Icon-56}. We have plotted the order parameter
$\left|\left< q_{k+1} - q_{k} \right> \right|$, the absolute values
of the ensemble average of adjacent particles relative displacement
(from their equilibrium positions), as a function of temperature $T$
for several space size $L$ (see figure 1) and verified that, for a
potential like equation (2), regardless of $L$, there is a phase
transition region around $T \simeq 0.02$-$0.1$. Second, we would
like to note that this system does not bear linear-wave dynamics
(with the unusual phonon dispersion)~\cite{DW-1}, then whether the
linear phonon dynamics would still apply should be further tested.
We expect that such two unusual features may affect heat transport
and to understand how they may is thus interesting.

\section{Simulation methods}
To derive the space-time scaling for heat transport, usually one can
focus on the dynamical exponent $\gamma$, defined by a
space($x$)-time($t$) scaling analysis of the system's heat spreading
density $\rho (x,t)$, i.e., $t^{-1/\gamma} \rho (t^{-1/\gamma}
x,t)$. This scaling exponent has been well predicted to follow
several universality classes by a recent celebrated theory of
nonlinear fluctuating
hydrodynamics~\cite{Hydrodynamics-1,Spohn2014-1,Spohn2014-2,Spohn2014-3}.
For a particular 1D system with conserved momentum and even
symmetric potential, $\gamma=3/2$ is suggested. From the L\'{e}vy
walks theory~\cite{Levy}, we may also have a formula $\alpha = 2
-\gamma$ that relates $\gamma$ to the space scaling exponent
$\alpha$. Therefore, besides $\alpha$, $\gamma$ can also be employed
to characterize the heat transport behavior, which however could
provide a more detailed information.

We here aim at identifying $\gamma$ from a direct dynamics
simulation. For such purpose, one can investigate the decay of
energy pulses or the equilibrium energy fluctuations
correlation~\cite{Diffusion2003,Diffusion2005,Diffusion2006zhao,Diffusion2011-2,Diffusion2013-1,Diffusion2013-2,Diffusion2013-3shun}.
However, studying the energy pluses decay may be unable to avoid
huge statistical fluctuations~\cite{Levy}, hence, here we apply the
equilibrium correlation method for our investigations. This
correlation approach was first proposed by
Zhao~\cite{Diffusion2006zhao} for studying the total energy
fluctuations spread and then extended to be applicable to
investigate both heat and other physical quantities fluctuations
decay~\cite{Diffusion2013-3shun}. For further detailed
implementation, one can also refer to~\cite{PingHuang}.

We shall mainly focus on the following three normalized
spatiotemporal correlation functions of the three main physical
quantities fluctuations, i.e., the heat energy, the total energy and
the momentum, defined as follows~\cite{Diffusion2006zhao,
Diffusion2013-3shun}
\begin{equation}
\rho_{Q} (x,t)=\frac{\langle \Delta Q_{j}(t) \Delta Q_{i}(0)
\rangle}{\langle \Delta Q_{i}(0) \Delta Q_{i}(0) \rangle};
\end{equation}
\begin{equation}
\rho_{E} (x,t)=\frac{\langle \Delta E_{j}(t) \Delta E_{i}(0)
\rangle}{\langle \Delta E_{i}(0) \Delta E_{i}(0) \rangle};
\end{equation}
\begin{equation}
\rho_{p}(x,t)=\frac{\langle \Delta p_{j}(t) \Delta p_{i}(0)
\rangle}{\langle \Delta p_{i}(0) \Delta p_{i}(0) \rangle},
\end{equation}
where $\langle \cdot \rangle$ represents the spatiotemporal average;
$\Delta Q_{i}(t)\equiv Q_i(t)- \langle Q_i \rangle$, $\Delta
E_{i}(t)\equiv E_i(t)- \langle E_i \rangle$, $\Delta p_{i}(t)\equiv
p_i(t)- \langle p_i \rangle$ are the corresponding fluctuations;
$Q_i(t) \equiv \sum Q(x,t)$, $E_i(t) \equiv \sum E(x,t)$ and $
p_i(t) \equiv \sum p(x,t)$ denote the heat energy, the total energy
and the momentum densities in an equal and appropriate lattice bin
$i$ (the number of particles in the $i$-th bin is equal to
$N_i=L/b$, where $b$ is the total number of the bins). For any
particle in each bin, $E(x,t)$ and $ p(x,t)$ are the
single-particle's total energy and momentum at the absolute
displacement $x$ and time $t$; $Q(x,t)\equiv E(x,t)-\frac{(\langle E
\rangle +\langle F \rangle)M(x,t)}{\langle M \rangle}$~\cite{Liquid}
is the particle's heat energy, with $M(x,t)$ the corresponding mass
density function, $\langle E \rangle$ ($\langle M \rangle$) and
$\langle F \rangle$ the spatiotemporally averaged energy (mass)
density and the internal pressure of the system in equilibrium
state, respectively.

Note that from the definition of $Q(x,t)$, it cannot be described as
a function of the lattice site, hence in practice we should have to
discretize the space into several bins, thus the space variable
should be the absolute displacement $x$ rather than the label $k$ of
the particle. We emphasize that the density of $\rho_{Q} (x,t)$
obtained from such a key coarse-grained procedure has been verified
to be more directly related to heat transport than the usually
considered site-site correlation of the total energy
fluctuations~\cite{Diffusion2013-3shun,Chenshunda-2}.

The correlation functions $\rho_{Q} (x,t)$, $\rho_{E} (x,t)$ and
$\rho_{p} (x,t)$ are employed to characterize the corresponding
spatiotemporal spreading processes of the initial physical
quantities fluctuations. If they have been obtained, then a scaling
analysis may enable us to identify the corresponding transport
manners.

As to our simulations, we assume the number of particles equal to
the space size $L$, then in view of the symmetric potential of the
system, the averaged pressure $\langle F \rangle$ is fixed at zero
throughout the simulations. We mainly consider two cases of space
size $L=2000$ and $ 4000$, which enables us to obtain an effective
space size (about $L_{\rm{effctive}}=500-1000$) for a long time up
to $t =200-600$ for the spread. For each $L$, we apply the periodic
boundary conditions, fix the bins number $b \equiv L/2$, and set the
lattice constant $a \equiv 1$ (the choice of $a$ has been verified
not to affect the final results). We utilize the stochastic Langevin
heat baths~\cite{Report-1,Report-2} to thermalize the system for
preparing the canonical equilibrium systems (with fixed $T$) and
employ the Runge-Kutta algorithm of $7$th to $8$th order with a time
step $0.05$ to evolve the system. The equilibrium systems are
prepared by evolving the systems for a long enough time ($>10^7$
time units of the models) from properly assigned initial random
states, then all the systems are evolved in isolation for deriving
the correlation information. The size of the ensemble for deriving
the correlations is about $8 \times 10^9$.
\section{Results}
\subsection{Heat transport}
Now let us first see the results of heat spread. Figure 2 depicts
the profiles of $\rho_Q(x,t)$ at a typical long time $t=600$ for
four temperatures, from low to high. In view of the transition
region shown in figure 1, the lowest temperature considered
(throughout the paper) is fixed at $T=0.02$, up to which we have
verified that the final results are insensitive to the initial
states assigned to just one of the potential wells; or between the
two wells alternately, or randomly. From figure 2 it can be clearly
seen that the shapes of the profiles under different temperatures
are different: while for a low temperature $T=0.02$ and a high
temperature $T=2.5$ one can identify one central peak and two side
peaks; in some intermediate temperature ranges, such as $T=0.05$ and
$T=0.1$, the side peaks seem to disappear. Thus, the latter cases of
$T=0.05$ and $T=0.1$ look like Gaussian distributions that are
usually exhibited in normal transport; while for the cases that the
side peaks do not disappear, $\rho_Q(x,t)$ are in good coincidence
with L\'{e}vy walks stable distributions for describing
superdiffusive transport~\cite{Levy}.
\begin{figure}
\vskip-.2cm \hskip-0.4cm \centering
\includegraphics[width=10.cm]{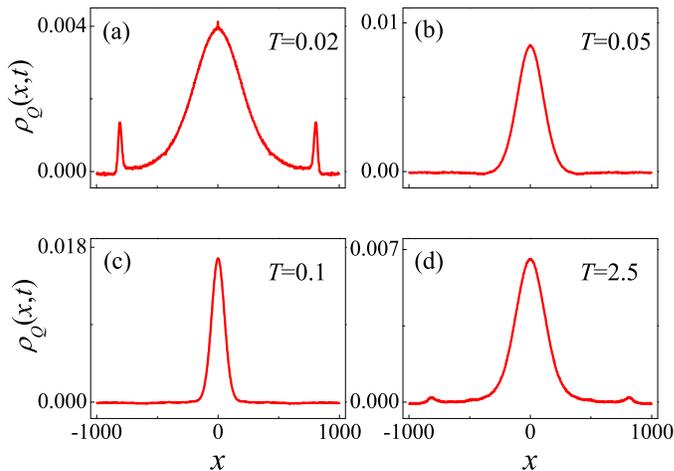}
\vskip-0.3cm \caption{(Color online) $\rho_Q(x,t)$ for time $t=600$
($L_{\rm{effective}}=2000$) under temperatures $T=0.02$ (a);
$T=0.05$ (b); $T=0.1$ (c) and $T=2.5$ (d), respectively.}
\end{figure}

Viewing this coincidence, we then perform a space-time scaling
analysis for the profiles central part
\begin{equation} \rho_{Q} (x,t) \simeq
\frac{1}{t^{1/\gamma}} \rho_{Q} (\frac{x}{t^{1/\gamma}},t),
\end{equation}
which then enables us to identify a dynamical scaling exponent
$\gamma$ for precisely characterizing the heat transport process. We
recall that usually $\gamma=1$, $1<\gamma<2$ and $\gamma=2$
correspond to the ballistic, superdiffusive, and normal transport,
respectively; and in particular, $\gamma=3/2$ is predicted by
nonlinear fluctuating
hydrodynamics~\cite{Hydrodynamics-1,Spohn2014-1,Spohn2014-2,Spohn2014-3}
for systems with symmetric potentials; $\gamma=5/3$ is found by both
L\'{e}vy walks approach and some numerical
simulations~\cite{Levy,Diffusion2003,Diffusion2005}.
\begin{figure}
\vskip-.2cm \hskip-0.4cm \centering
\includegraphics[width=10.cm]{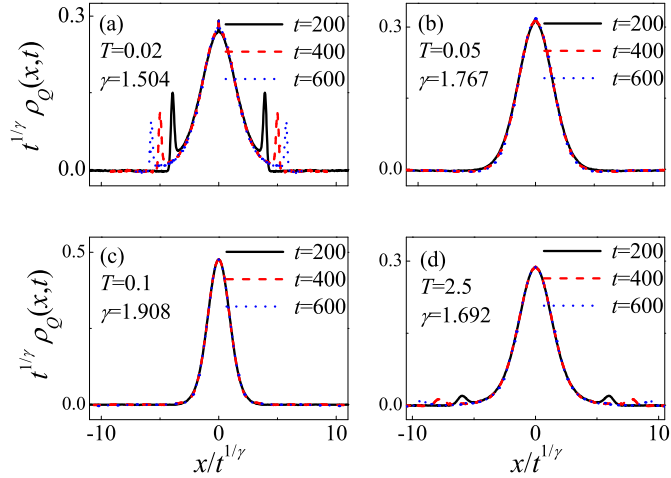}
\vskip-0.3cm \caption{(Color online) Rescaled $\rho_{Q}(x,t)$ shown
in figure 2: (a) $T=0.02$ ($\gamma=1.504$); (b) $T=0.05$
($\gamma=1.767$); (c) $T=0.1$ ($\gamma=1.908$); and (d) $T=2.5$
($\gamma=1.692$), respectively. In each curve scaling $t^{1/\gamma}
\rho (x,t)$ vs. $x/t^{1/\gamma}$ for three different times are
compared.}
\end{figure}

After obtaining the scaling exponent from formula (6), in figure 3
we plot the rescaled $\rho_{Q}(x,t)$ for four typical temperatures
considered in figure 2. As can be seen, for all of the temperatures,
formula (6) is beautifully satisfied suggesting that the focused
system's heat spread can be well captured by the single-particle's
L\'{e}vy walk models; though the scaling exponent $\gamma$ is
different for different temperatures: (i) for the lowest temperature
$T=0.02$, $\gamma \simeq 1.504$, in good agreement with the
prediction $\gamma=3/2$ of nonlinear fluctuating hydrodynamics for
even symmetric
potentials~\cite{Hydrodynamics-1,Spohn2014-1,Spohn2014-2,Spohn2014-3};
(ii) in the case of high temperature $T=2.5$, the best fitting gives
$\gamma \simeq 1.692$, consistent well with the popular numerical
results of $\gamma=5/3$ based on the L\'{e}vy walks
models~\cite{Levy,Diffusion2003,Diffusion2005,Diffusion2011-2,Diffusion2013-1,Diffusion2013-2};
(iii) while in the intermediate ranges of temperature, $\gamma$
tends to increase; in particular, around $T \simeq 0.1$, $\gamma
\simeq 1.908$, suggesting that a heat transport process very close
to normal ($\gamma = 2$) diffusion appears to take place, which
coincides well with the early numerical findings of normal heat
conduction ($\alpha \simeq 0$) in this temperature
regimes~\cite{Report-1,Rotator-1,DW}.

Then what are the pictures for other temperatures? To answer this
question, we carefully examine the results of $\gamma(T)$ and
summarize them in figure 4. Therein four data points are extracted
from figure 3, while others are calculated (fitted) additionally in
the same way. For each $T$, two long effective space size of
$L_{\rm{effective}}=1000$ and $2000$ have been considered for the
analysis of finite size effects. From figure 4 one can see that
regardless of $L$, as $T$ increases from $T=0.02$ to $T=2.5$,
$\gamma$ increases first from $\gamma \simeq 3/2$, reaches its
maximum value close to $\gamma=2$ at $T_{\rm{tr}} \simeq 0.1$, then
decreases down to $\gamma \simeq 5/3$ for $T=2.5$. Thus, $\gamma$
appears not a universal constant for DW systems, independence of
$T$, and a crossover from superdiffusive ($1<\gamma<2$) to normal
transport ($\gamma=2$) at about $T_{\rm{tr}} \simeq 0.1$ is likely
to take place, though longer space sizes simulations are still
required to confirm the transition. This result is consistent with
the results of $\alpha$ reported in~\cite{DW}, but obviously more
precise and detailed.
\begin{figure}
\vskip-.2cm \hskip-0.4cm \centering
\includegraphics[width=10cm]{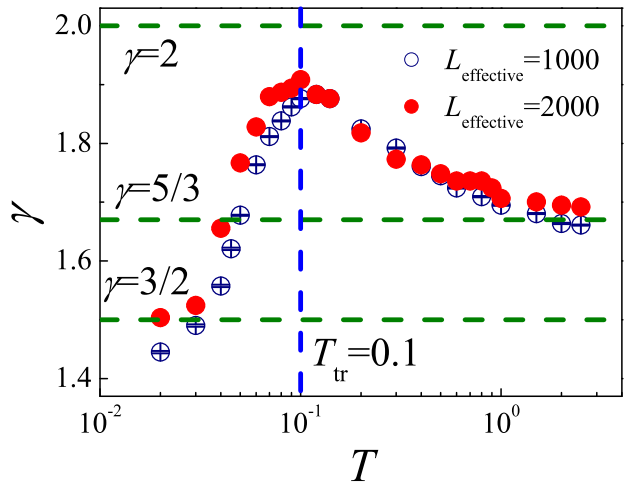}
\vskip-0.3cm \caption{(Color online) $\gamma$ vs. $T$, where the
hollow (solid) circle corresponds to $L_{\rm{effective}}=1000$
($2000$), and the horizontal dashed lines, from bottom to top,
denote $\gamma=3/2$, $\gamma=5/3$ and $\gamma=2$; the vertical
dashed line denotes $T_{\rm{tr}}=0.1$, respectively.}
\end{figure}

\subsection{Transport of total energy}
\begin{figure}
\vskip-.2cm \hskip-0.4cm \centering
\includegraphics[width=10cm]{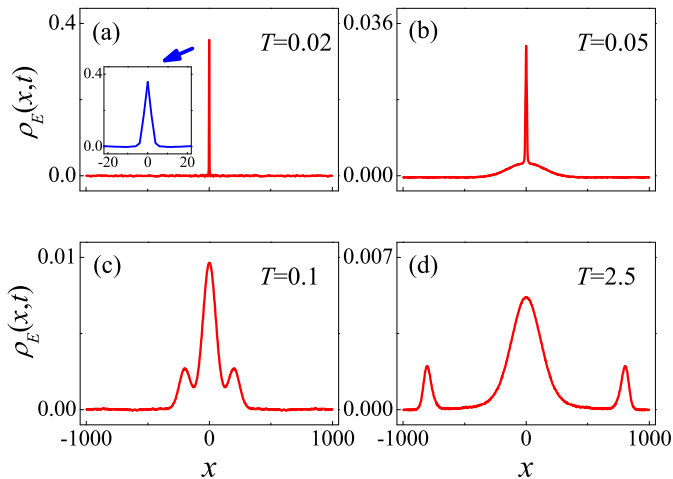}
\vskip-0.3cm \caption{(Color online) $\rho_{E} (x,t)$ for time
$t=600$ ($L_{\rm{effective}}=2000$) under system temperatures
$T=0.02$ (a); $T=0.05$ (b); $T=0.1$ (c) and $T=2.5$ (d),
respectively.}
\end{figure}
Next, we turn to the results of the total energy spread. Our focus
will be limited to demonstrating the distinctions between the total
energy diffusion and heat spread. We will show that the spreading
processes of these two physical quantities could be very different,
thus one should take very care to relate heat conduction to just
energy diffusion~\cite{ShaLiu-1}.

Figure 5 shows $\rho_{E} (x,t)$ for four temperatures, the same as
those considered in figure 2, which then enables us to make a quick
comparison of the heat and the total energy spread. From figure 5 it
can be seen that, despite that in the high temperatures, $\rho_{E}
(x,t)$ shows one central peak and two side peaks as well [see figure
5(d)], similarly to $\rho_{Q} (x,t)$; for the cases of low
temperatures, $\rho_{E} (x,t)$ implies very strong localizations
[see figure 5(a)]; while around the crossover temperature point of
$T \simeq 0.1$, there is a transition from localization to
delocalization for energy [see figure 5(b)-(c)]. This phenomenon is
very strange, since under some temperature regimes, the total energy
is strongly localized, while heat can be superdiffusive [see figure
2(a)], the mechanisms are thus interesting and we wish to understand
via further studies.

Finally, we would like to note that though the difference between
$\rho_{Q} (x,t)$ and $\rho_{E} (x,t)$ is slight in the frequently
considered FPU-$\beta$ systems (under certain appropriate
temperature regimes)~\cite{Diffusion2013-3shun}, and there previous
attempt to relate heat conduction to just the total energy diffusion
in fact does not deviate too much; however, in the case of 1D DW
systems considered here, concerning the heat spread obviously seems
more reasonable.
\subsection{Momentum transport}
\begin{figure}
\vskip-.2cm \hskip-0.4cm \centering
\includegraphics[width=10cm]{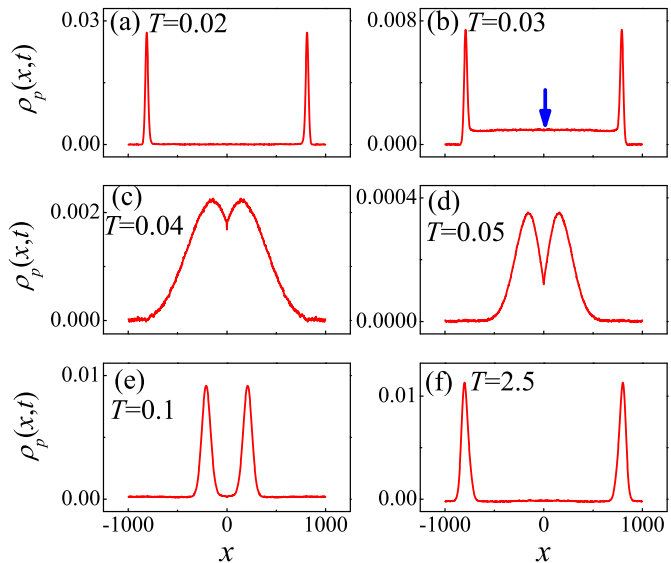}
\vskip-0.4cm \vskip-0.3cm \caption{(Color online) $\rho_{p} (x,t)$
for time $t=600$ ($L_{\rm{effective}}=2000$) under system
temperatures $T=0.02$ (a); $T=0.03$ (b); $T=0.04$ (c); $T=0.05$ (d);
$T=0.1$ (e); and $T=2.5$ (f), respectively.}
\end{figure}
In order to further explore the mechanisms of normal heat transport,
we then move on to the spreading process of the third physical
quantity, the momentum, since a quite recent work has attributed the
normal heat transport to the diffusive behavior of momentum
spread~\cite{YunyunLi}.

Figure 6 depicts $\rho_{p}(x,t)$ at time $t=600$ for six
temperatures, among which, four of them, i.e., $T=0.02$, $T=0.05$,
$T=0.1$ and $T=2.5$ are those considered in figures 2-3 and 5, while
two additional temperatures $T=0.03$ and $T=0.04$ are added for
further demonstrating the details of transition. From figure 6 it
can be seen that, in both cases of low and high temperatures, there
are ballistic type spreading of momentum [see figure 6(a) and (f)];
however, in the intermediate temperature ranges, the non-ballistic,
non-Gaussian behaviors can be clearly identified [see figure
6(b)-(e)].

The feature of momentum spread might be characterized by:
\begin{equation}
\langle \Delta x_{p}^2 \left( t \right) \rangle = \sum_{x} x^2
\rho_{p}(x,t) \sim t^{\mu},
\end{equation}
where $\langle \Delta x_{p}^2 \left( t \right) \rangle$ is the mean
squared deviation (MSD) of momentum spread, $\mu$ is its time
scaling exponent. Then $\mu=2$ corresponds to ballistic spreading of
momentum; while for $\mu=1$, $\rho_{p}(x,t)$ may turn to a Gaussian
distribution, thus implying a diffusive behavior of momentum spread,
which has been conjectured to be related directly to the normal heat
transport in the 1D rotator system~\cite{YunyunLi}. Finally,
$1<\mu<2$ then lies somewhere in between, suggesting the
superdiffusive behavior.

\begin{figure}
\vskip-.2cm \hskip-0.4cm \centering
\includegraphics[width=10cm]{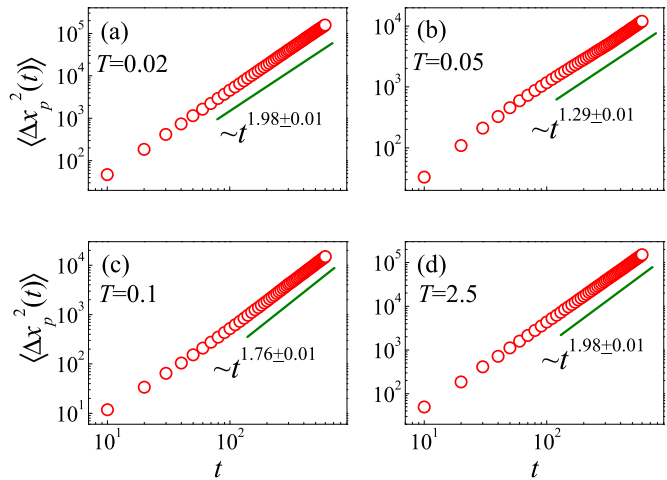}
\vskip-0.3cm \caption{(Color online) $\langle \Delta x_{p}^2 \left(
t \right) \rangle$ of momentum spread vs. $t$ for temperatures
$T=0.02$ ($\mu=1.98 \pm 0.01$) (a); $T=0.05$ ($\mu=1.29 \pm 0.01$)
(b); $T=0.1$ ($\mu=1.76 \pm 0.01$) (c), and $T=2.5$ ($\mu=1.98 \pm
0.01$) (d), respectively.}
\end{figure}

Figure 7 presents some typical results of the MSD $\langle \Delta
x_{p}^2 \left( t \right) \rangle$ vs. $t$ (log-log), a linear
fitting then gives the scaling exponent $\mu$. Indeed, in both cases
of $T=0.02$ and $T=2.5$, our best fittings suggest that $\mu$ are
very close to $2$, thus supporting the ballistic spreading process
of momentum; while for $T=0.05$ and $T=0.1$, the best fittings show
$\mu \simeq 1.29$ and $\mu \simeq 1.76$, implying the non-ballistic,
non-Gaussian superdiffusive behavior. Now it is clear that if the
momentum spread with different $\mu$ can actually be related to heat
transport, then the $\mu$ of $T=0.05$ obviously shows less value
than that of $T=0.1$. We thus further examine $\mu$ as a function of
$T$ and summarize the result in figure 8. As expected, a crossover
from ballistic to non-ballistic momentum spread is appearing to take
place; while it is $T_{\rm{tr}}=0.05$, rather than $T_{\rm{tr}}=0.1$
being the turning point. Unfortunately, this turning point of
momentum spread seems not directly related to the crossover point of
$T_{\rm{tr}}=0.1$ found in normal transport of heat, though it may
play a role.
\begin{figure}
\vskip-.2cm \hskip-0.4cm \centering
\includegraphics[width=10cm]{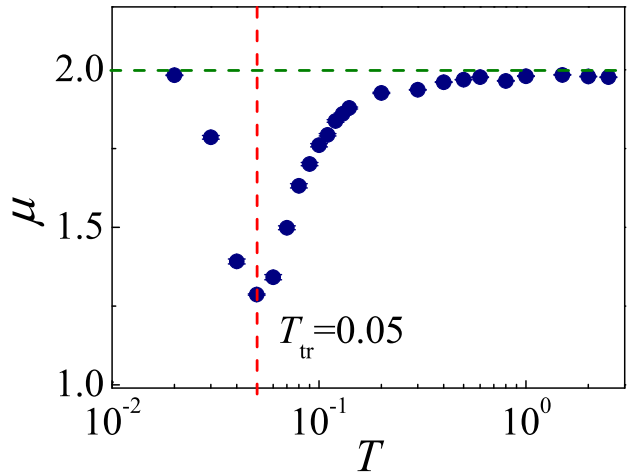}
\vskip-0.3cm \caption{(Color online) Scaling exponent $\mu$ of
momentum spread vs. $T$, where the vertical (horizontal) line
denotes $T_{\rm{tr}}=0.05$ ($\mu=2$).}
\end{figure}

In fact, such an unusual feature of momentum spread can also be
detected from other measurements. For example, it is interesting to
reveal the scaling properties of the peaks shown in $\rho_{p}(x,t)$.
For such purpose one may examine how the height $h$ of the peaks
decays with time $t$. Since it is reasonable to assume that the
peaks of $\rho_{p}(x,t)$ keep their volumes unchanged over the time,
such an examination actually enables us to explore the dispersion of
the peaks and so gain the information of sound attenuation. We have
verified that the height $h$ will scale with $t$ as
$t^{-\lambda}$~\cite{Diffusion2013-3shun} in the considered long
time, so here we just plot the exponent $\lambda$ as a function of
temperature $T$ in figure 9. As can be seen, the result also
suggests the turning point of $T_{\rm{tr}}=0.05$, coincident well
with the measurement of $\mu$. Another detail is that in the high
temperature regimes, the exponent $\lambda$ seems to finally
converge to $\lambda=0.5$, which just gives the previous numerical
findings of the exponent in the FPU-$\beta$
chains~\cite{Diffusion2013-3shun}.
\begin{figure}
\vskip-.2cm \hskip-0.4cm \centering
\includegraphics[width=10cm]{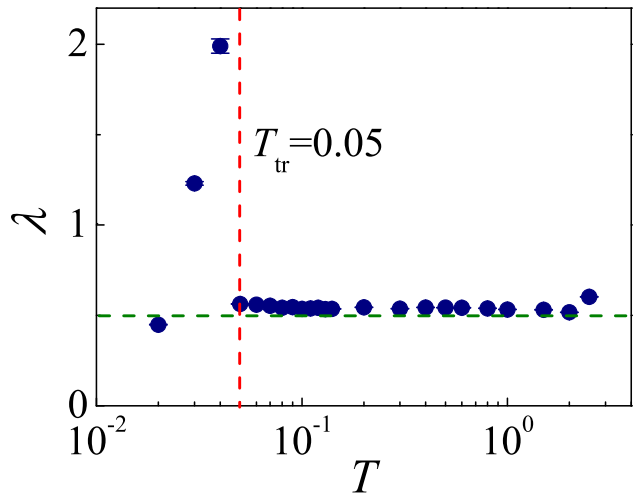}
\vskip-0.3cm \caption{(Color online) Scaling exponent $\lambda$ of
the peaks in momentum spread vs. $T$, where the vertical
(horizontal) line denotes $T_{\rm{tr}}=0.05$ ($\lambda=0.5$).}
\end{figure}
\begin{figure}
\vskip-.2cm \hskip-0.4cm \centering
\includegraphics[width=10cm]{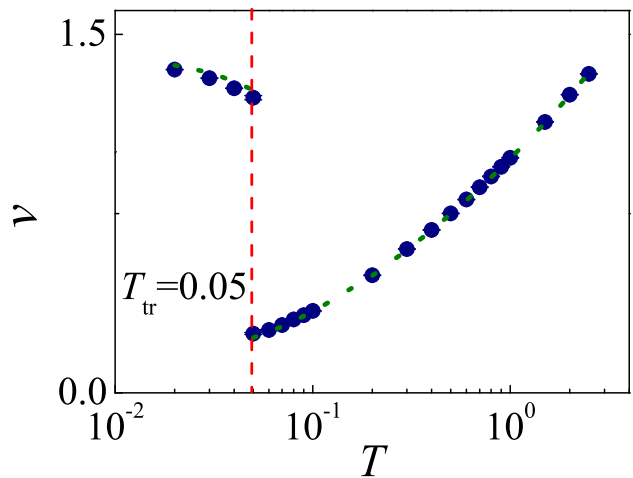}
\vskip-0.3cm \caption{(Color online) The sound velocity $v$ vs. $T$,
where the vertical dashed line denotes $T_{\rm{tr}}=0.05$; the
dotted lines are the predictions from~\cite{Spohn2014-1}.}
\end{figure}

It is also interesting to measure the velocity $v$ of the peaks
(which is usually suggested to correspond to the sound velocity) and
compare it with the recent predictions in nonlinear fluctuating
hydrodynamics~\cite{Spohn2014-1}. Figure 10 provides such a result
of $v$ vs. $T$. The predictions are from the formula addressed
in~\cite{Spohn2014-1}
\begin{equation}
v=\sqrt{ \frac{\frac{1}{2} T^{2} + \left\langle V + \left\langle
F\right\rangle \xi; V + \left\langle F\right\rangle \xi \right
\rangle } { \frac{1}{T} \left( \left\langle \xi; \xi \right \rangle
\left<V; V \right \rangle - \left\langle \xi; V \right\rangle ^{2}
\right)
 +\frac{1}{2} T \left\langle \xi; \xi \right\rangle } },
\end{equation}
where $V(\xi)$ is the potential, $\left\langle A; B \right\rangle$
denotes the covariance $\left\langle A B \right\rangle -
\left\langle A \right\rangle \left\langle B \right\rangle$ for any
two quantities $A$ and $B$, and $\left\langle F \right\rangle$ is
the averaged pressure ($\equiv 0$ in this case). To obtain the
predictions, usually one can insert the DW potential into equation
(8) and calculate the ensemble average of each quantity
$\left\langle A \right\rangle$ by $\int_{-\infty}^{\infty} A
e^{-V(\xi)/T} \rm{d} \xi / \int_{-\infty}^{\infty}
e^{-\mit{V}(\xi)/\mit{T}} \rm{d} \xi$. While for the temperatures
blow the turning point of $T_{\rm{tr}}=0.05$, it is reasonable to
consider the integrations only over $\xi \geq 0$, i.e.,
$\left\langle A \right\rangle = \int_{0}^{\infty} A e^{-V(\xi)/T}
\rm{d} \xi / \int_{0}^{\infty} e^{-\mit{V}(\xi)/\mit{T}} \rm{d}
\xi$, since in view of the properties of the DW potential, under low
temperatures, only one of the wells can be covered. So for just
$T_{\rm{tr}}=0.05$, we provide both predictions from different ways
of integrations. In fact, from the simulations, around the turning
point of $T_{\rm{tr}}=0.05$, we can indeed identify two different
ballistic peaks with different velocities in some relative short
time (see figure 11). Now from figure 10, one can see that the
velocities match well with the predictions, suggesting that the
predictions of nonlinear fluctuating hydrodynamics can also be
validated to the 1D DW systems; and more-importantly, the result
clearly indicates the turning point of $T_{\rm{tr}}=0.05$.
\begin{figure}
\vskip-.2cm \hskip-0.4cm \centering
\includegraphics[width=10cm]{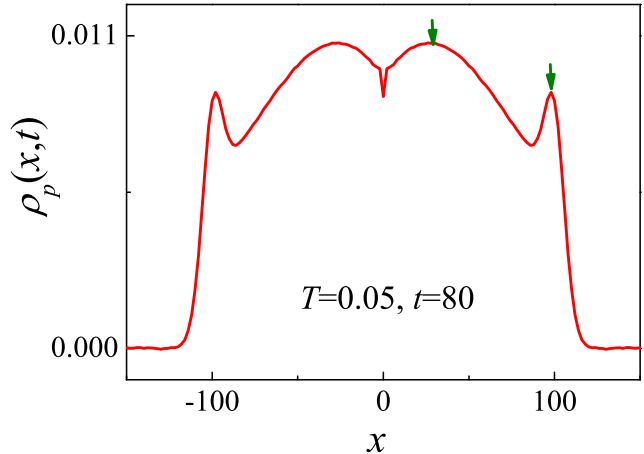}
\vskip-0.3cm \caption{(Color online) $\rho_{p} (x,t)$ for a relative
short time $t=80$ at the turning point of temperature
$T_{\rm{tr}}=0.05$.}
\end{figure}

\subsection{Phonons spectra}
The above two crossover (turning) temperature points for heat and
momentum spread naturally puzzle us. In order to better understand
the mechanisms, we finally turn to analyzing how the phonons spectra
$P(\omega)$ of this system would depend on temperature, from which
we may gain some suggestive information.

Figure 12 depicts $P(\omega)$ vs. $T$ for six typical temperatures
considered in figure 6. For each temperature, $P(\omega)$ is
calculated by applying a frequency $\omega$ analysis of the
equilibrium states particles velocity $v(t)$ along the systems,
i.e., $P(\omega)=\lim_{\tau \rightarrow \infty} \frac{1}{\tau}
\int_{0}^{\tau} v (t) \exp (- \rm{i} \mit{\omega} t) \rm{d}
\mit{t}$. From figure 12 it can been seen that $P(\omega)$ also
shows strong dependence of $T$, especially that in some temperature
ranges, phonons tend to become ``softer". These tendencies can be
readily captured from the denoted peaks in the high frequency
regimes.
\begin{figure}
\vskip-.2cm \hskip-0.4cm \centering
\includegraphics[width=10cm]{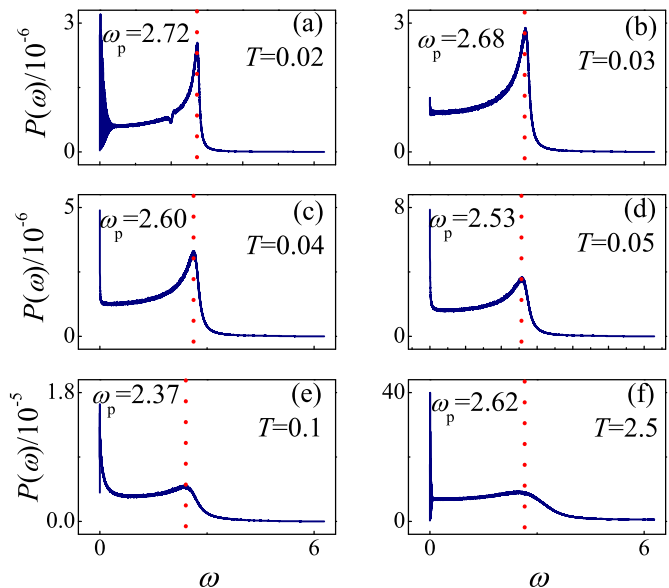}
\vskip-0.4cm \vskip-0.3cm \caption{(Color online) Power spectra
$P(\omega)$ vs. $T$: (a) $T=0.02$; (b) $T=0.03$; (c) $T=0.04$; (d)
$T=0.05$; (e) $T=0.1$; and (f) $T=2.5$. The dotted lines denote the
peaks with frequency $\omega_{\rm{p}}$ in the high frequency
regimes.}
\end{figure}

To characterize the phonons softening and to see how it is related
to thermal transport, we plot the averaged frequency $\bar{\omega}$,
defined by $\bar{\omega}={\int_{0}^{\infty} P (\omega) \omega \rm{d}
\mit{\omega}} / {\int_{0}^{\infty} P (\omega) \rm{d} \mit{\omega}}$,
as a function of $T$ in figure 13. As can be seen, besides the
turning point of $T_{\rm{tr}} \simeq 0.1$ ($T_{\rm{tr}} \simeq
0.05$) for heat (momentum) spread, there is another turning point of
$T_{\rm{tr}} \simeq 0.2$ for phonons softening. Given that
$T_{\rm{tr}} \simeq 0.1$ just lies somewhere in between, we may
conjecture that the observed normal heat transport is probably
induced by the combined effects of phonons softening and the
non-ballistic momentum spread.
\begin{figure}
\vskip-.2cm \hskip-0.4cm \centering
\includegraphics[width=10cm]{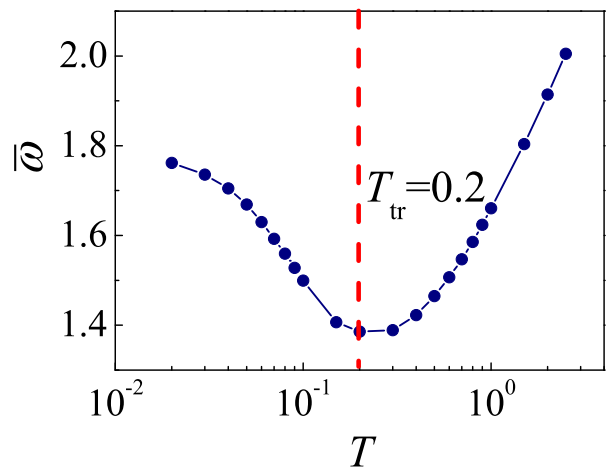}
\vskip-0.3cm \caption{(Color online) The averaged frequency
$\bar{\omega}$ from $ P (\omega)$ vs. $T$, where the dashed line
denotes $T_{\rm{tr}}=0.2$.}
\end{figure}

About the phonon spectra calculated here, we would also like to
point out that the non-ballistic behavior of momentum spread can
also be detected from the phonon's lowest frequency components. We
address this point by using a log-log plot of figure 12 (see figure
14). Phonons with the lowest frequency, usually called
long-wavelength (goldstone) modes, are generally believed to be very
weakly damped due to the conserved feature of
momentum~\cite{Newbook}. Because of their weak damping, the lowest
frequency modes can greatly affect heat transport. From figure 14,
it can be clearly seen that with the increase of $T$, the damping of
phonons first originates from the high frequencies and then quickly
walks towards the low ones. Surprisingly one may identify that a
complete damping appears around the turning point of $T_{\rm{tr}}
\simeq 0.05$ [see Figure 14(d)], the same as that found in momentum
spread. This result clearly indicates a strong positive correlation
between the phonons damping and the non-ballistic spreading of
momentum, thus implying that the origin of the long-wavelength modes
may be induced by the ballistic spreading of momentum.
\begin{figure}
\vskip-.2cm \hskip-0.4cm \centering
\includegraphics[width=10cm]{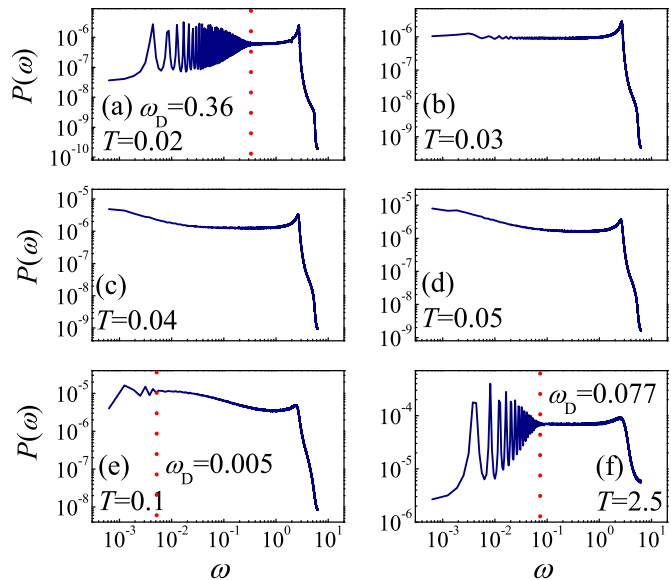}
\vskip-0.4cm \vskip-0.3cm \caption{(Color online) Log-log plot of
figure 12, where the dotted lines denote the frequencies
$\omega_{\rm{D}}$ below which phonons are damped very weakly.}
\end{figure}
\section{Conclusions}
To summarize, we have studied in detail the rich heat transport
behaviors in the focused 1D anharmonic oscillator systems with a DW
interparticle interaction like $V(\xi)=-\xi^2/2+\xi^4/4$. By
employing the equilibrium correlation method, we have captured the
profiles of heat spread and precisely identified its space-time
scaling exponents under various system temperatures. These heat
spreading profiles are in good coincidence with the L\'{e}vy walks
stable distributions, and their scaling laws show good satisfactory
with L\'{e}vy walks scaling as well. Based on the scaling, we are
able to present the precise temperature dependence of heat transport
in this system and further verify that there is a crossover from
superdiffusive to normal heat transport with a turning point at
about $T_{\rm{tr}} \simeq 0.1$ in this particular system. This
result thus provides more detailed and precise numerical evidences
clearly demonstrating that normal heat transport is likely to take
place in the 1D systems with DW interactions under the appropriate
temperature regimes, though the total momentum is conserved here.

In order to explore the mechanisms of the observed normal heat
transport, we have carefully examined the spreading of two other
physical quantities, i.e., the total energy and momentum. The
spreading of the total energy is found to be very distinct from heat
spread. In particular, under some temperature regimes, heat can be
superdiffusive, while energy shows strong localizations. This
unusual result thus suggests that we should take care to derive a
general connection between heat conduction and energy diffusion;
rather, it may be reasonable to connect heat conduction to heat
spread.

The momentum spread is shown to have a second crossover from
ballistic to non-ballistic behaviors with a second turning point of
$T_{\rm{tr}} \simeq 0.05$; however, this turning point of momentum
spread cannot directly correspond to the first turning point of heat
spread ($T_{\rm{tr}} \simeq 0.1$). So to understand the observed
normal heat transport, taking only the non-ballistic behavior of
momentum spread into account is inadequate. We then perform an
analysis of the phonons spectra of the system. We find that phonons
tend to become softest around another turning point of $T_{\rm{tr}}
\simeq 0.2$. Together with the $T_{\rm{tr}} \simeq 0.05$ of momentum
spread, we thus conjecture that it is the combined effects of the
non-ballistic momentum spread and the phonons softening that result
in the normal heat transport observed in this system. All of the
simulation results seem not to contradict this conjecture, though
further detailed investigations remain required.

Finally, we would like to point out that, once we have understood
the mechanisms, then apart from this theoretical advance, there is
also room for possible applications. For example, one may be able to
vary the phonons spectrum by adjusting temperatures in other DW
systems, and finally manipulate heat. Such an idea would be realized
by variation of the trapping frequencies in the recent focused ion
chains~\cite{Icon-1234}. This kind of systems has been found to
exhibit a structural phase transition similar to the DW
systems~\cite{Icon-56}, with which then heat transport could be
tunable.

\ack  The author would like to thank the referee for his many
helpful suggestions, and Profs. Hong Zhao, Jiao Wang and Yong Zhang
from Xiamen university of China for their huge valuable discussions.
This work was supported by the National Natural Science Foundation
of China (Grants No. 11575046 and No. 11205032), the Natural Science
Foundation of Fujian province, China (No. 2013J05008).

\end{document}